\documentclass[aps,twocolumn,amssymb]{revtex4}
\usepackage{amsmath,amssymb,amsfonts}
\usepackage{graphics}

\begin{document}
\title{EXAFS and XRD studies of an amorphous 
Co$_{57}$Ti$_{43}$ alloy produced by mechanical alloying}

\author{K. D. Machado}
\email{kleber@fisica.ufsc.br}
\affiliation{Departamento de F\'{\i}sica, Universidade Federal de Santa Catarina, 88040-900 
Florian\'opolis, SC, Brazil}

\author{J. C. de Lima}
\affiliation{Departamento de F\'{\i}sica, Universidade Federal de Santa Catarina, 88040-900 
Florian\'opolis, SC, Brazil}

\author{C. E. M. Campos}
\affiliation{Departamento de F\'{\i}sica, Universidade Federal de Santa Catarina, 88040-900 
Florian\'opolis, SC, Brazil}

\author{T. A. Grandi}
\affiliation{Departamento de F\'{\i}sica, Universidade Federal de Santa Catarina, 88040-900 
Florian\'opolis, SC, Brazil}

\author{A. A. M. Gasperini}
\affiliation{Departamento de F\'{\i}sica, Universidade Federal de Santa Catarina, 88040-900 
Florian\'opolis, SC, Brazil}

\date{\today}
%\pacs{61.10.Eq}{x-ray scattering}
%\pacs{61.10.Ht}{EXAFS technique}
%\pacs{61.43.Dq}{Amorphous alloys}
%\pacs{61.43.Bn}{structural modeling} 
%\pacs{81.05.Kf}{Metallic glasses}

\begin{abstract}
We have investigated the local atomic structure of an amorphous Co$_{57}$Ti$_{43}$ alloy produced by 
Mechanical Alloying by means of x-ray diffraction and EXAFS analyses on Co and Ti K-edges. Coordination 
numbers and 
interatomic distances where found and compared with those determined using an additive hard sphere 
(AHS) model associated with a RDF$(r)$ deconvolution, and also with data from 
bcc-Co$_2$Ti compound. The EXAFS results obtained indicated a shortening in the Co-Ti and Ti-Ti 
distances when compared to those found by the AHS-RDF method and an increase in the Co-Co 
and Ti-Ti distances and a large shortening in the Co-Ti one when compared to the distances found 
in the bcc-Co$_2$Ti compound. In 
spite of these differences, coordination numbers obtained from EXAFS and AHS-RDF are similar to 
each other and also to those found in bcc-Co$_2$Ti.

\end{abstract}


\maketitle

\section{Introduction}

Mechanical alloying (MA) technique \cite{MA} is an efficient means for synthesizing 
crystalline and amorphous materials, as well as stable and metastable solid 
solutions \cite{KleSeZn,KleberNiTi,Dolgin,Hellstern,Kimura,Yavari}. MA has also been 
used to produce materials with nanometer sized grains and alloys whose components have large 
differences in their melting temperatures and are thus difficult to produce using techniques based on 
melting. The few thermodynamics restrictions on the alloy composition open up a wide range of 
possibilities for property combinations even for immiscible elements 
\cite{Abbate}. Since temperatures reached in MA are usually very low, this low temperature process 
reduces reaction kinetics, allowing 
the production of poorly crystallized or amorphous materials. 
Recently, we have used MA to produce amorphous Ni$_{60}$Ti$_{40}$ \cite{KleberNiTi} and 
partially amorphous Fe$_{60}$Ti$_{40}$ 
\cite{JoaoFeTi} alloys. Ni-Ti alloys are biocompatible and can be used to produced shape memory 
materials, and Fe-Ti alloys can be used as hydrogen storage materials. It is known that various 
cobalt-rich Co-TM (TM = transition metals) amorphous 
alloys obtained by sputtering can provide good soft magnetic characteristics, such as high saturation 
magnetization and low coercivity. However, sputtering produces only thin amorphous films \cite{Naka}, 
and melt-spinning has some problems in preparing amorphous ribbon and powders \cite{Nose}. Thus, 
here we have used MA to produce an amorphous Co$_{57}$Ti$_{43}$ alloy ({\em a}-Co$_{57}$Ti$_{43}$) 
starting from the crystalline elemental powders. Its structural properties were studied by x-ray 
diffraction (XRD) and Extended X-ray absorption fine structure (EXAFS) techniques. 
Due to its selectivity and high sensitivity to the chemical environment around a specific 
type of atom of an alloy, EXAFS \cite{Teo,Lee,Hayes,Rehrrev,Prins} is a technique 
very suitable to investigate the local atomic order of crystalline compounds and amorphous alloys. 
Anomalous wide angle x-ray scattering (AWAXS) is also a selective technique, but due to the small 
$K_{\rm{max}}$ that can be achieved on Ti K-edge ($\sim 4$ \AA$^{-1}$), little information could 
be obtained from an AWAXS experiment on this edge. On the other hand, EXAFS measurements 
performed on Ti edge extended to $\sim 15$ \AA$^{-1}$. Using this technique, we have determined 
coordination numbers and interatomic distances in the first coordination shell of 
{\em a}-Co$_{57}$Ti$_{43}$. In addition, we have modelled the atomic structure of 
{\em a}-Co$_{57}$Ti$_{43}$ by using an additive hard sphere (AHS) model (in which the minimum distance 
between unlike atoms is given by $D_{12}=(D_{11}+D_{22})/2$, where $D_{ij}$ is the closest approach 
distance between the centers of atoms $i$ and $j$), 
to perform the deconvolution of the radial distribution functions 
(RDF$(r)$). The coordination numbers and interatomic distances obtained from the AHS-RDF model were 
compared to those found from EXAFS, and the EXAFS results obtained indicated a shortening in the Co-Ti 
and Ti-Ti distances when compared to those found by the AHS-RDF method. We have also compared the 
structural parameters of the alloy with those found in the bcc-Co$_2$Ti compound. In 
spite of some differences, coordination numbers obtained from EXAFS and AHS-RDF are similar to 
each other and also to those found in bcc-Co$_2$Ti.

\section{Faber-Ziman structure factors}

According to Faber and Ziman \cite{Faber}, the total structure factor ${\cal S}(K)$ is obtained 
from the scattered intensity per atom $I_a(K)$ through

\begin{eqnarray}
{\cal S}(K) &=& \frac{I_a(K)-\bigl[\langle f^2(K)\rangle - \langle f(K)\rangle^2
\bigr]}{\langle f(K)\rangle^2} \label{eqstructurefactor1} \\
&= & \sum_{i=1}^n{\sum_{j=1}^n{w_{ij}(K) {\cal S}_{ij}(K) }}\,,
\label{eqstructurefactor}
\end{eqnarray}

\noindent where $K$ is the transferred momentum,  
${\cal S}_{ij}(K)$ are the partial structure factors and 

\begin{equation}
w_{ij}(K) = \frac{c_i c_j f_i(K) f_j(K)}{\langle f(K)\rangle^2}\,,
\end{equation}

\noindent and

\begin{eqnarray}
\langle f^2(K) \rangle &=& \sum_{i}{ c_i f_i^2(K)}\,, \\
\langle f(K) \rangle^2 &=& \Bigl[\sum_{i}{ c_i f_i(K)}\Bigr]^2 \,.
\label{eqw}
\end{eqnarray}

\noindent Here, $f_i(K)$ is the atomic scattering factor 
and $c_i$ is the concentration of atoms of type $i$. The partial and total pair 
distribution functions $G_{ij}(r)$ and $G(r)$ are related to ${\cal S}_{ij}(K)$ and 
${\cal S}(K)$ through

\begin{eqnarray}
G_{ij}(r) &=& \frac{2}{\pi} \int_0^{\infty}{K\bigl[{\cal S}_{ij}(K)-1 \bigr] 
\sin (Kr)\, dK}\,,\\
G(r) &=& \frac{2}{\pi} \int_0^{\infty}{K\bigl[{\cal S}(K)-1 \bigr] 
\sin (Kr)\, dK}\,.
\end{eqnarray}

\noindent From the $G_{ij}(r)$ and $G(r)$ functions the partial and total radial distribution 
functions RDF$_{ij}(r)$ and RDF$(r)$ can be calculated by

\begin{eqnarray}
{\rm{RDF}}_{ij}(r) &=& 4\pi \rho_0 c_j r^2+ r G_{ij}(r)\,, \\
{\rm{RDF}}(r) &=& 4\pi \rho_0 r^2+ r G(r)\,,
\label{trdf}
\end{eqnarray}

\noindent where $\rho_0$ is the density of the alloy (in atoms/\AA$^3$). Interatomic distances 
are obtained from the maxima of $G_{ij}(r)$ and 
coordination numbers are calculated by integrating the peaks of RDF$_{ij}(r)$. 

\section{Additive Hard Sphere Model}
\label{secAHS}

In the AHS model for a binary alloy, the interacting potentials between particles $i$ and $j$ are 

\begin{equation}
u_{ij} =
\begin{cases}
0, & r > D_{ij}\\
\infty, & r < D_{ij}
\end{cases}
\end{equation}

\noindent where $D_{ij}$ is the closest approach distance between the centers of the particles $i$ 
and $j$, and 

\begin{equation}
D_{12} = \frac{D_{11}+D_{22}}{2} \,,
\end{equation}

\noindent that is, the minimum approach between unlike atoms is always equal to the arithmetic mean of 
the diameters $D_{11}$ and $D_{22}$ of the two species. For this model, the Percus-Yevick (PY) equation, 
which can be applied to systems in which short range forces are dominant, has an exact solution, making 
it possible to obtain analytical expressions for the partial pair distribution functions and partial 
structure factors. In 1977, Weeks \cite{Weeks} used the PY model to study the atomic structure of 
metallic glasses, assuming an isotropic and homogeneous phase that does not occur in the solid state. 
The extension of the PY model to study the glassy state is based on the structural characteristics of 
amorphous state, which can be considered as an extrapolation of the atomic structure of the liquid state. 
A good review of the AHS model is given in Ref. \cite{Waseda}.

\section{Experimental Procedure}

Blended Co (Vetec, 99.7\%, particle size $< 10$ $\mu$m) and Ti (BDH, 99.5\%, particle size $< 10$ 
$\mu$m) elemental powders, with nominal composition Co$_{60}$Ti$_{40}$, were 
sealed together with several steel balls (with diameter of about 1 cm), under an argon atmosphere, in 
a steel vial. The 
ball-to-powder weight ratio was 5:1. The vial was mounted in a high energy ball mill Spex Mixer/Mill 
model 8000 (working at 1200 rpm) and milled for 
10 h. A ventilation system was used to keep the vial temperature close to room temperature. The 
composition of the as-milled powder was measured using the Energy Dispersive Spectroscopy (EDS) 
technique, giving a composition of 57 at. \% and 43 at. \% of Co and Ti, respectively, and impurity 
traces were not observed. 
XRD measurements were recorded on a Siemens diffractometer with a graphite monochromator in the 
diffracted beam, using the Cu K$_\alpha$ line ($\lambda = 1.5418$ \AA). ${\cal S}(K)$ was computed 
from the XRD patterns, after standard corrections following the procedure 
described by Wagner \cite{Wagner}.
The samples for the EXAFS measurements were formed by placing the powder onto a porous menbrane 
(Millipore, 0.2 $\mu$m pore size) in order to achive optimal thickness (about 30 $\mu$m), and neither 
Kapton tape nor BN were used. The 
EXAFS measurements were taken at room temperature in the transmission mode on the 
D04B beam line of LNLS (Campinas, Brazil), using a channel cut monochromator (Si 111) and two 
ionization chambers filled by air as detectors, working at 10\% and 70\% efficiency, respectively, 
and the beam size at the sample was about 3 $\times$ 1 mm. This yielded a resolution of about 2.0 eV 
on Ti K edge and 2.5 eV on Co K edge, respectively. At these energies, harmonic rejection is irrelevant 
at the D04B beam line. 
The energy and average current of the storage ring were 1.37 GeV and 120 mA, respectively.

\section{Results and Discussion}

\begin{figure}
\includegraphics{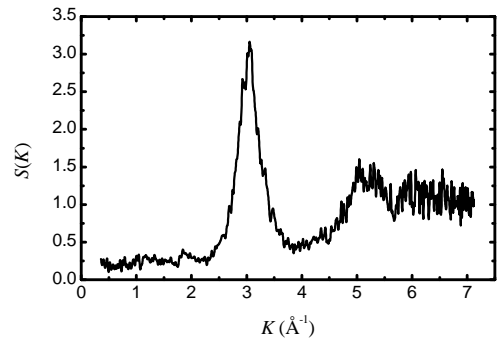}
\caption{Experimental total structure factor for {\em a}-Co$_{57}$Ti$_{43}$.}
\label{fig1sk}
\end{figure}

\begin{figure}
\includegraphics{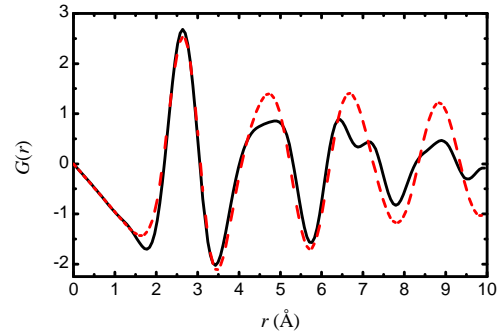}
\caption{Experimental (solid line) and AHS (dashed line) $G(r)$ functions for 
{\em a}-Co$_{57}$Ti$_{43}$.}
\label{fig2g}
\end{figure}

\begin{figure}
\includegraphics{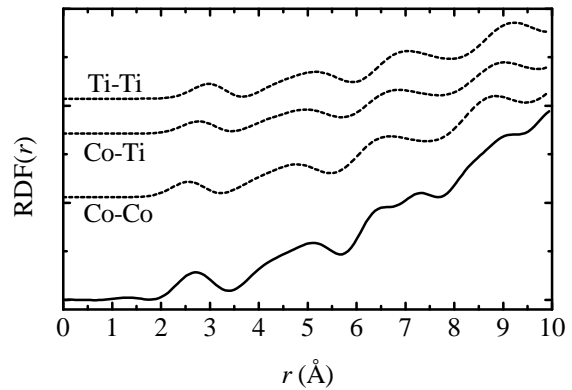}
\caption{Experimental RDF$(r)$ (solid line) and partial RDF$^{\rm AHS}_{ij}(r)$ 
(dashed lines) functions for {\em a}-Co$_{57}$Ti$_{43}$.}
\label{fig3rdf}
\end{figure}

Figure \ref{fig1sk} shows ${\cal S}(K)$ for {\em a}-Co$_{57}$Ti$_{43}$. 
In this figure a diffuse halo between $K = 2.2$ and 3.8 \AA$^{-1}$ can be seen, indicating the presence of an 
amorphous phase. Residual crystalline peaks of the elemental metal powders are not observed. 
The glass-forming ability of the Co-Ti system made by MA was investigated by 
Dolgin {\em et al} \cite{Dolgin}, Hellstern and Schultz 
\cite{Hellstern} and Kimura {\em et al} \cite{Kimura}. They reported amorphization for this system, 
in good agreement with our results.
The local atomic structure of {\em a}-Co$_{57}$Ti$_{43}$ was firstly studied considering an AHS model. 
We used the AHS model to obtain ${\cal S}^{\rm AHS}_{ij}(K)$ and 
$G^{\rm AHS}_{ij}(r)$ for a system with the same composition of the alloy. It is well known that the 
intensity of the main halo of the total structure factor ${\cal S}^{\rm AHS}(K)$ generated by the AHS 
model is larger than the experimental one. Thus, ${\cal S}^{\rm AHS}(K)$ has to be multiplied by a 
$\exp{(- \sigma^2K^2)}$ function in order to introduce a `thermal' effect. The best agreement between the 
experimental $G(r)$ and $G^{\rm AHS}(r)$ functions were achieved when the packing fraction parameter and 
$\sigma^2$ values were 0.76 and 0.07 \AA$^2$, respectively. Figure \ref{fig2g} shows $G(r)$ (full line) 
and $G^{\rm AHS}(r)$ (dashed line) obtained for {\em a}-Co$_{57}$Ti$_{43}$. There is an excellent 
agreement concerning peak positions, in particular the first one, but there are some differences in 
the intensities of the peaks starting from the second one, which could be explained by the features of 
the AHS model. It is interesting to note that the AHS model was developed to investigate atomic structures 
of glassy alloys in the liquid state and, as the chemical short-range order (CSRO) of the alloy becomes 
stronger, it is not able to reproduce the structural features of the alloy anymore. Thus, the differences 
reported above can be associated with the presence of a CSRO in {\em a}-Co$_{57}$Ti$_{43}$. This CSRO 
probably is not much strong since the differences between experimental and AHS data are not very 
large. The density of the amorphous alloy can be calculated from the slope of the straight line 
($-4\pi \rho_0 r$) fitting the initial part (until the first minimum) of the $G(r)$ function 
\cite{Waseda}. We have found $\rho_0 = \rho^{\rm AHS} = 0.0761$ 
atoms/\AA$^3$ for both experimental and AHS $G(r)$ functions, whereas the average density 
is $\langle \rho \rangle = c_{\rm Co} \rho_{\rm Co}+c_{\rm Ti} \rho_{\rm  Ti} = 0.0763$ 
atoms/\AA$^3$. 
Figure \ref{fig3rdf} shows the experimental RDF$(r)$ and the 
partial RDF$^{\rm AHS}_{ij}(r)$ (see eq. \ref{trdf}) functions obtained from the AHS model. From 
these functions, coordination numbers and interatomic distances between first neighbors can be found, 
and they are given in table \ref{tab1}. 
It should be noted that due to the small values of the 
weighting factor $w_{\rm Ti-Ti}(K)$ ($w_{\rm Ti-Ti}(K) \approx 10$\%, see eq. \ref{eqw}), it is very 
difficult to obtain reliable structural data concerning Ti-Ti pairs using the AHS-RDF method.

\begin{table}
\caption{\label{tab1} Structural data determined for {\em a}-Co$_{57}$Ti$_{43}$.}
\begin{center}
\begin{tabular}{cccccc}\hline
\multicolumn{6}{c}{EXAFS} \\\hline
& \multicolumn{2}{c}{Co K-edge} & \multicolumn{3}{c}{Ti K-edge}\\\hline
Bond Type & Co-Co & Co-Ti & Ti-Co & \multicolumn{2}{c}{Ti-Ti\footnotemark}\\
$N$ & 6.0  & 6.0 & 7.9  & 1.9 & 3.0 \\
$r$ (\AA) & 2.50 & 2.52 & 2.52 & 2.81 & 3.08\\
$\sigma^2$ (\AA $\times 10^{-2}$) & 1.45 & 4.57 & 4.57& 1.46 & 1.29  \\\hline
\multicolumn{6}{c}{AHS-RDF} \\\hline
Bond Type & Co-Co & Co-Ti & Ti-Co & \multicolumn{2}{c}{Ti-Ti}\\
$N$ & 6.5 & 5.5 & 7.3 & \multicolumn{2}{c}{4.0} \\
$r$ (\AA) & 2.49 & 2.75 & 2.75 & \multicolumn{2}{c}{3.17}\\\hline
\multicolumn{6}{c}{bcc-Co$_2$Ti \cite{TAPP}} \\\hline
Bond Type & Co-Co & Co-Ti & Ti-Co & \multicolumn{2}{c}{Ti-Ti}\\
$N$ & 6 & 6 & 12 & \multicolumn{2}{c}{4} \\
$r$ (\AA) & 2.37 & 2.77 & 2.77 & \multicolumn{2}{c}{2.90}\\\hline
\end{tabular}
\centerline{($^1$) There are 4.9 Ti-Ti pairs at $\langle r \rangle = 2.96$ \AA.}
\end{center}
\end{table}

\begin{figure}
\includegraphics{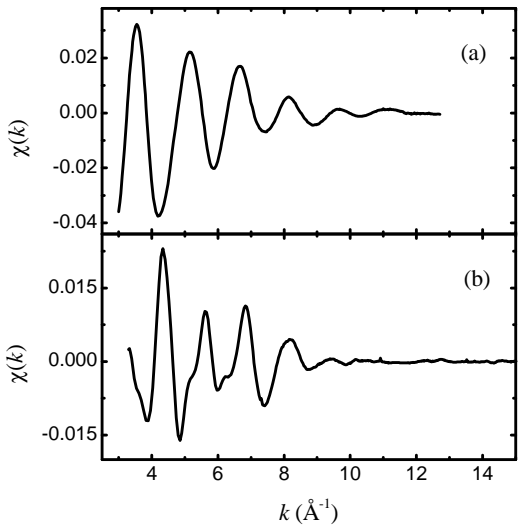}
\caption{Experimental EXAFS spectra: (a) Co K edge and (b) Ti K edge.}
\label{fig1}
\end{figure}

\begin{figure}
\includegraphics{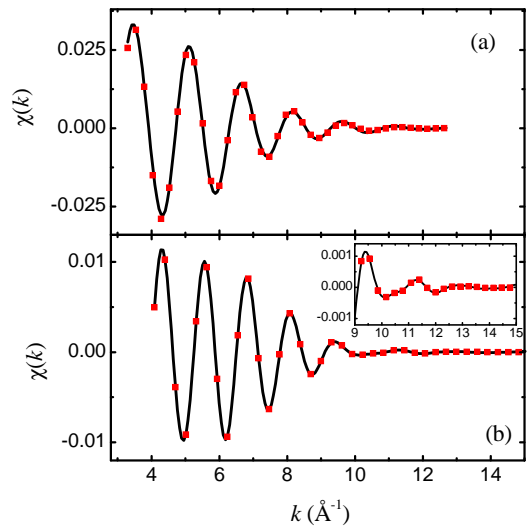}
\caption{Fourier-filtered first shell (full line) and its simulation (squares) 
on (a) Co K edge, (b) Ti K edge. The inset at (b) shows the high-$k$ 
EXAFS data and its simulation.}
\label{fig3}
\end{figure}

The EXAFS oscillations $\chi(k)$ at both K edges are shown in fig.~\ref{fig1}. After standard data 
reduction procedures using Winxas97 software \cite{Ressler}, they were filtered by Fourier transforming 
$k^3\chi(k)$ (Co edge, 3.00 -- 12.71 \AA$^{-1}$) and $k\chi(k)$ (Ti edge, 3.32 -- 15.00 \AA$^{-1}$) 
using a Hanning weighting function into $r$-space and transforming back the 
first coordination shells (1.30 -- 2.67 \AA\ for Co edge and 1.85 -- 3.24 \AA\ for Ti edge). Filtered 
spectra were then fit by using Gaussian distributions to represent the homopolar and heteropolar 
bonds \cite{Stern}. We also used the third cumulant option of Winxas97 to investigate the presence of 
asymmetric shells. The amplitude and phase shifts relative to the homopolar and heteropolar bonds 
needed to fit them were obtained from {\em ab initio} calculations using the spherical waves method 
\cite{Rehr} and FEFF software.
Figure~\ref{fig3} shows the experimental and the fitting results for the Fourier-filtered first 
shells on Co and Ti edges. The high quality of the fit on Ti edge is evidenced on the inset at  
fig. \ref{fig3}.b, which shows the high-$k$ EXAFS data on this edge together with its simulation. 
Structural parameters extracted from the fits are listed in 
table~\ref{tab1}.
As it can be seen in this table, on Co edge one Co-Co shell and one Co-Ti shell were 
considered in the first shell in order to find a good fit, whereas on Ti edge one 
Ti-Cu shell and two Ti-Ti subshells  were needed. We started the fitting procedure using one 
shell for all pairs. This choice did not produce a good quality fit of Fourier-filtered first shells 
on the Ti edge. Besides that, the well-known relations 
$c_i N_{ij} = c_j N_{ji}$, $r_{ij} = r_{ji}$ and $\sigma_{ij} = \sigma_{ji}$,
where $c_i$ is the concentration of atoms of type $i$, $N_{ij}$ is the number of $j$ atoms located at 
a distance $r_{ij}$ around an $i$ atom and $\sigma_{ij}$ is the half-width of the Gaussian, 
were not verified in fitting EXAFS data. By 
considering two Ti-Ti subshells the quality of the fit on the Ti edge was much improved (see 
fig.~\ref{fig3}). Moreover, the relations above were satisfied. It should be noted that Nyquist 
criterion as defined in ref. \cite{Stern2} is also satisfied even with two Ti-Ti subshells due to the 
large $k$-data range.

Comparing EXAFS and AHS-RDF results it can be seen that concerning coordination numbers they show 
a good agreement, mainly if the difficulties in finding data about Ti-Ti pairs by the AHS-RDF analysis 
mentioned above are considered. In addition, coordination numbers determined by EXAFS have an error 
of about 1 atom for this kind of alloys. However, except for the Co-Co interatomic distance, AHS-RDF 
results overestimate distance values, which could be explained by the considerations made in the 
definition of the AHS model. This decrease in the distances is not unexpected because, according 
to Hausleitner and Hafner \cite{Hausleitner}, which investigated several amorphous alloys formed by 
transition 
metals (TM-Ti) using molecular dynamics simulations, obtaining structure factors, coordination numbers 
and interatomic distances, if the components of an alloy 
have a large difference in the number of $d$ electrons there is a pronounced non-additivity of the 
pair interactions and a strong interaction between unlike atoms, thus decreasing the distance between 
chemically different atoms. The shortening effect is even stronger in {\em a}-Ni$_{60}$Ti$_{40}$ 
\cite{KleberNiTi} produced by 
MA. This alloy has 8.8 Ni-Ni pairs at $\langle r \rangle = 2.58$ \AA, 5.2 Ni-Ti pairs 
at $\langle r \rangle = 2.55$ \AA\ and 5.5 Ti-Ti pairs at $\langle r \rangle = 2.90$ \AA, and in 
this case Ni-Ti pairs are even shorter than Ni-Ni pairs. Comparing {\em a}-Ni$_{60}$Ti$_{40}$ 
and {\em a}-Co$_{57}$Ti$_{43}$ it can be seen that structural data are similar, the main difference 
being the number of TM-TM pairs (TM = Ni or Co), which is larger in the first alloy probably because 
of its stronger CSRO. 
In addition, the shortening effect was also found by Fukunaga {\em et al}. in 
{\em a}-Ni$_{40}$Ti$_{60}$ produced by melt-quenching \cite{Fukunaga}, and recently 
we have also seen it in {\em a}-Cu$_{64}$Ti$_{36}$ produced by MA, whose data will be published 
elsewhere (we have chosen not to 
compare with Fe$_{60}$Ti$_{40}$ \cite{JoaoFeTi} because this alloy is 
not completely amorphous). 
It is interesting to note that the local structure of {\em a}-Co$_{57}$Ti$_{43}$ shows some 
differences when compared to that found in the bcc-Co$_2$Ti compound. The compositional difference 
may explain the disagreement between Co-Ti coordination numbers, but there is an important reduction 
in the Co-Ti distance in the amorphous alloy, whereas Co-Co and Ti-Ti distances increase. 
These features usually are attributed to the preparation technique, that introduces many defects and 
vacancies in the alloy, and they have also been 
reported for other alloys produced by MA \cite{KleberNiTi,KleberGeSe,KleberGaSe}.

In order to estimate the CSRO in the alloy we used the generalized Warren parameter 
$\alpha_{\rm W}$ given by \cite{Gazzillo}

\begin{equation}
\alpha_{\rm W} = 1 - \frac{N_{12}}{c_2 N} = 1 - \frac{N_{21}}{c_1 N}\nonumber\,,
\end{equation}

\noindent where $N_{ij}$ are the coordination numbers and $N_i = \sum_{j}{N_{ij}}$ and 
$N = c_1N_2 + c_2 N_1$.  
The $\alpha_{\rm W}$ parameter is null for a random distribution. If there is a preference for forming 
unlike pairs in the alloy, it becomes negative. Otherwise, it is positive if homopolar pairs are preferred. 
Here, using the coordination numbers given in table \ref{tab1}, we found $\alpha_{\rm W}^{\rm EXAFS} 
= -0.117$ and $\alpha_{\rm W}^{\rm AHS} = -0.103$. Since the values of $\alpha_{\rm W}$ are 
negative and small, they indicate a weak CSRO in the alloy, in agreement with the results above.

\section{Conclusion}

An amorphous Co$_{57}$Ti$_{43}$ alloy was produced by Mechanical Alloying technique and its local 
atomic structure was obtained from EXAFS analysis, which furnished coordination numbers and 
interatomic distances between first neighbors. The results found were compared with 
those determined by using an additive hard sphere model combined with an RDF$(r)$ deconvolution 
process. Concerning coordination numbers, there is a good agreement between EXAFS and AHS-RDF 
data. However, AHS-RDF furnishes interatomic distances for Co-Ti and Ti-Ti pairs too large when 
compared to EXAFS results, which could be attributed to the way the AHS model is defined. In 
addition, we have also seen the decrease in Co-Ti distance as proposed by Hausleitner and Hafner 
\cite{Hausleitner} for TM-Ti alloys. There is a small CSRO in the alloy, and a preference for 
forming unlike pairs. The local structure of {\em a}-Co$_{57}$Ti$_{43}$ is not much different 
from that found in {\em a}-Ni$_{60}$Ti$_{40}$ except for the number of TM-TM pairs (TM = Ni or Co), 
which is larger in the second alloy. However, {\em a}-Co$_{57}$Ti$_{43}$ has a 
local structure different from that found in bcc-Co$_2$Ti compound, as seen in 
other alloys produced by MA \cite{KleberNiTi,KleberGeSe,KleberGaSe}. 

\acknowledgments
We thank CNPq, CAPES and LNLS (proposal no XAS 799/01) for financial support.

%\bibliographystyle{apsrev}
%\bibliography{co57ti43}

\begin{thebibliography}{30}
\expandafter\ifx\csname natexlab\endcsname\relax\def\natexlab#1{#1}\fi
\expandafter\ifx\csname bibnamefont\endcsname\relax
  \def\bibnamefont#1{#1}\fi
\expandafter\ifx\csname bibfnamefont\endcsname\relax
  \def\bibfnamefont#1{#1}\fi
\expandafter\ifx\csname citenamefont\endcsname\relax
  \def\citenamefont#1{#1}\fi
\expandafter\ifx\csname url\endcsname\relax
  \def\url#1{\texttt{#1}}\fi
\expandafter\ifx\csname urlprefix\endcsname\relax\def\urlprefix{URL }\fi
\providecommand{\bibinfo}[2]{#2}
\providecommand{\eprint}[2][]{\url{#2}}

\bibitem[{\citenamefont{Suryanarayana}(2001)}]{MA}
\bibinfo{author}{\bibfnamefont{C.}~\bibnamefont{Suryanarayana}},
  \bibinfo{journal}{Prog. Mater. Sci.} \textbf{\bibinfo{volume}{{46}}},
  \bibinfo{pages}{1} (\bibinfo{year}{2001}).

\bibitem[{\citenamefont{Machado et~al.}(2003)\citenamefont{Machado, de~Lima,
  Campos, Grandi, and Gasperini}}]{KleSeZn}
\bibinfo{author}{\bibfnamefont{K.~D.} \bibnamefont{Machado}},
  \bibinfo{author}{\bibfnamefont{J.~C.} \bibnamefont{de~Lima}},
  \bibinfo{author}{\bibfnamefont{C.~E.~M.} \bibnamefont{Campos}},
  \bibinfo{author}{\bibfnamefont{T.~A.} \bibnamefont{Grandi}},
  \bibnamefont{and} \bibinfo{author}{\bibfnamefont{A.~A.~M.}
  \bibnamefont{Gasperini}}, \bibinfo{journal}{Sol. State Commun.}
  \textbf{\bibinfo{volume}{127}}, \bibinfo{pages}{477} (\bibinfo{year}{2003}).

\bibitem[{\citenamefont{Machado et~al.}(2002)\citenamefont{Machado, de~Lima,
  de~Campos, Grandi, and Trich\^es}}]{KleberNiTi}
\bibinfo{author}{\bibfnamefont{K.~D.} \bibnamefont{Machado}},
  \bibinfo{author}{\bibfnamefont{J.~C.} \bibnamefont{de~Lima}},
  \bibinfo{author}{\bibfnamefont{C.~E.~M.} \bibnamefont{de~Campos}},
  \bibinfo{author}{\bibfnamefont{T.~A.} \bibnamefont{Grandi}},
  \bibnamefont{and} \bibinfo{author}{\bibfnamefont{D.~M.}
  \bibnamefont{Trich\^es}}, \bibinfo{journal}{Phys. Rev. B}
  \textbf{\bibinfo{volume}{66}}, \bibinfo{pages}{094205}
  (\bibinfo{year}{2002}).

\bibitem[{\citenamefont{Dolgin et~al.}(1986)\citenamefont{Dolgin, Vanek,
  McGory, and Ham}}]{Dolgin}
\bibinfo{author}{\bibfnamefont{B.~P.} \bibnamefont{Dolgin}},
  \bibinfo{author}{\bibfnamefont{M.~A.} \bibnamefont{Vanek}},
  \bibinfo{author}{\bibfnamefont{T.}~\bibnamefont{McGory}}, \bibnamefont{and}
  \bibinfo{author}{\bibfnamefont{D.~J.} \bibnamefont{Ham}},
  \bibinfo{journal}{J. Non-Cryst. Sol.} \textbf{\bibinfo{volume}{87}},
  \bibinfo{pages}{281} (\bibinfo{year}{1986}).

\bibitem[{\citenamefont{Hellstern and Schultz}(1987)}]{Hellstern}
\bibinfo{author}{\bibfnamefont{E.}~\bibnamefont{Hellstern}} \bibnamefont{and}
  \bibinfo{author}{\bibfnamefont{L.}~\bibnamefont{Schultz}},
  \bibinfo{journal}{Mater. Sci. and Eng.} \textbf{\bibinfo{volume}{93}},
  \bibinfo{pages}{213} (\bibinfo{year}{1987}).

\bibitem[{\citenamefont{Kimura et~al.}(1988)\citenamefont{Kimura, Takada, and
  W.-N.}}]{Kimura}
\bibinfo{author}{\bibfnamefont{H.}~\bibnamefont{Kimura}},
  \bibinfo{author}{\bibfnamefont{F.}~\bibnamefont{Takada}}, \bibnamefont{and}
  \bibinfo{author}{\bibfnamefont{W.-N.~M.} \bibnamefont{W.-N.}},
  \bibinfo{journal}{Mat. Sci. Eng.} \textbf{\bibinfo{volume}{97}},
  \bibinfo{pages}{125} (\bibinfo{year}{1988}).

\bibitem[{\citenamefont{Yavari et~al.}(1992)\citenamefont{Yavari, Desr\'e, and
  Benameur}}]{Yavari}
\bibinfo{author}{\bibfnamefont{A.~R.} \bibnamefont{Yavari}},
  \bibinfo{author}{\bibfnamefont{P.~J.} \bibnamefont{Desr\'e}},
  \bibnamefont{and} \bibinfo{author}{\bibfnamefont{T.}~\bibnamefont{Benameur}},
  \bibinfo{journal}{Phys. Rev. Lett.} \textbf{\bibinfo{volume}{{68}}},
  \bibinfo{pages}{2235} (\bibinfo{year}{1992}).

\bibitem[{\citenamefont{Abbate et~al.}(2001)\citenamefont{Abbate, Schreiner,
  Grandi, and de~Lima}}]{Abbate}
\bibinfo{author}{\bibfnamefont{M.}~\bibnamefont{Abbate}},
  \bibinfo{author}{\bibfnamefont{W.~H.} \bibnamefont{Schreiner}},
  \bibinfo{author}{\bibfnamefont{T.~A.} \bibnamefont{Grandi}},
  \bibnamefont{and} \bibinfo{author}{\bibfnamefont{J.~C.}
  \bibnamefont{de~Lima}}, \bibinfo{journal}{J. Phys.: Cond. Matter}
  \textbf{\bibinfo{volume}{13}}, \bibinfo{pages}{5723} (\bibinfo{year}{2001}).

\bibitem[{\citenamefont{Lima et~al.}(2003)\citenamefont{Lima, Machado, Drago,
  Grandi, Campos, and Trich\^es}}]{JoaoFeTi}
\bibinfo{author}{\bibfnamefont{J.~C.~D.} \bibnamefont{Lima}},
  \bibinfo{author}{\bibfnamefont{K.~D.} \bibnamefont{Machado}},
  \bibinfo{author}{\bibfnamefont{V.}~\bibnamefont{Drago}},
  \bibinfo{author}{\bibfnamefont{T.~A.} \bibnamefont{Grandi}},
  \bibinfo{author}{\bibfnamefont{C.~E.~M.} \bibnamefont{Campos}},
  \bibnamefont{and} \bibinfo{author}{\bibfnamefont{D.~M.}
  \bibnamefont{Trich\^es}}, \bibinfo{journal}{J. Non-Cryst. Solids}
  \textbf{\bibinfo{volume}{318}}, \bibinfo{pages}{121} (\bibinfo{year}{2003}).

\bibitem[{\citenamefont{Naka et~al.}(1984)\citenamefont{Naka, Kazama, Fujimori,
  and Masumoto}}]{Naka}
\bibinfo{author}{\bibfnamefont{M.}~\bibnamefont{Naka}},
  \bibinfo{author}{\bibfnamefont{N.~S.} \bibnamefont{Kazama}},
  \bibinfo{author}{\bibfnamefont{H.}~\bibnamefont{Fujimori}}, \bibnamefont{and}
  \bibinfo{author}{\bibfnamefont{T.}~\bibnamefont{Masumoto}},
  \emph{\bibinfo{title}{Proc. 5$^{\rm th}$ Int. Conf. Rapidly Quenched Metals}}
  (\bibinfo{publisher}{Wurzburg}, \bibinfo{year}{1984}).

\bibitem[{\citenamefont{Nose et~al.}(1982)\citenamefont{Nose, Esashi, Kanehira,
  Ohnuma, Shirakawa, and Masumoto}}]{Nose}
\bibinfo{author}{\bibfnamefont{M.}~\bibnamefont{Nose}},
  \bibinfo{author}{\bibfnamefont{K.}~\bibnamefont{Esashi}},
  \bibinfo{author}{\bibfnamefont{J.}~\bibnamefont{Kanehira}},
  \bibinfo{author}{\bibfnamefont{S.}~\bibnamefont{Ohnuma}},
  \bibinfo{author}{\bibfnamefont{K.}~\bibnamefont{Shirakawa}},
  \bibnamefont{and} \bibinfo{author}{\bibfnamefont{T.}~\bibnamefont{Masumoto}},
  \emph{\bibinfo{title}{Proc. 4$^{\rm th}$ Int. Conf. Rapidly Quenched Metals}}
  (\bibinfo{publisher}{Wurzburg}, \bibinfo{year}{1982}).

\bibitem[{\citenamefont{Teo and Joy}(1981)}]{Teo}
\bibinfo{author}{\bibfnamefont{B.~K.} \bibnamefont{Teo}} \bibnamefont{and}
  \bibinfo{author}{\bibfnamefont{D.~C.} \bibnamefont{Joy}},
  \emph{\bibinfo{title}{EXAFS Spectroscopy, Techniques and Applications}}
  (\bibinfo{publisher}{Plenum}, \bibinfo{address}{New York},
  \bibinfo{year}{1981}).

\bibitem[{\citenamefont{Lee et~al.}(1981)\citenamefont{Lee, Citrin,
  Eisenberger, and Kincaid}}]{Lee}
\bibinfo{author}{\bibfnamefont{P.~A.} \bibnamefont{Lee}},
  \bibinfo{author}{\bibfnamefont{P.}~\bibnamefont{Citrin}},
  \bibinfo{author}{\bibfnamefont{P.}~\bibnamefont{Eisenberger}},
  \bibnamefont{and} \bibinfo{author}{\bibfnamefont{B.}~\bibnamefont{Kincaid}},
  \bibinfo{journal}{Rev. Mod. Phys.} \textbf{\bibinfo{volume}{53}},
  \bibinfo{pages}{769} (\bibinfo{year}{1981}).

\bibitem[{\citenamefont{Hayes and Boyce}(1982)}]{Hayes}
\bibinfo{author}{\bibfnamefont{T.~M.} \bibnamefont{Hayes}} \bibnamefont{and}
  \bibinfo{author}{\bibfnamefont{J.~B.} \bibnamefont{Boyce}},
  \emph{\bibinfo{title}{Solid State Physics}} (\bibinfo{publisher}{Academic
  Press}, \bibinfo{address}{New York}, \bibinfo{year}{1982}),
  vol.~\bibinfo{volume}{37}, p. \bibinfo{pages}{173}.

\bibitem[{\citenamefont{Rehr and Albers}(2000)}]{Rehrrev}
\bibinfo{author}{\bibfnamefont{J.~J.} \bibnamefont{Rehr}} \bibnamefont{and}
  \bibinfo{author}{\bibfnamefont{R.~C.} \bibnamefont{Albers}},
  \bibinfo{journal}{Rev. Mod. Phys.} \textbf{\bibinfo{volume}{72}},
  \bibinfo{pages}{621} (\bibinfo{year}{2000}).

\bibitem[{\citenamefont{Koningsberger and Prins}(1988)}]{Prins}
\bibinfo{author}{\bibfnamefont{D.~C.} \bibnamefont{Koningsberger}}
  \bibnamefont{and} \bibinfo{author}{\bibfnamefont{R.}~\bibnamefont{Prins}},
  \emph{\bibinfo{title}{X-ray Aabsorption}} (\bibinfo{publisher}{Wiley},
  \bibinfo{address}{New York}, \bibinfo{year}{1988}).

\bibitem[{\citenamefont{Faber and Ziman}(1965)}]{Faber}
\bibinfo{author}{\bibfnamefont{T.~E.} \bibnamefont{Faber}} \bibnamefont{and}
  \bibinfo{author}{\bibfnamefont{J.~M.} \bibnamefont{Ziman}},
  \bibinfo{journal}{Philos. Mag.} \textbf{\bibinfo{volume}{11}},
  \bibinfo{pages}{153} (\bibinfo{year}{1965}).

\bibitem[{\citenamefont{Weeks}(1977)}]{Weeks}
\bibinfo{author}{\bibfnamefont{J.~D.} \bibnamefont{Weeks}},
  \bibinfo{journal}{Phil. Mag.} \textbf{\bibinfo{volume}{{35}}},
  \bibinfo{pages}{1345} (\bibinfo{year}{1977}).

\bibitem[{\citenamefont{Waseda}(1980)}]{Waseda}
\bibinfo{author}{\bibfnamefont{Y.}~\bibnamefont{Waseda}},
  \emph{\bibinfo{title}{The Structure of Non-Crystalline Materials (Liquid and
  Amorphous Solids)}} (\bibinfo{publisher}{McGraw-Hill}, \bibinfo{address}{New
  York}, \bibinfo{year}{1980}).

\bibitem[{\citenamefont{Wagner}(1972)}]{Wagner}
\bibinfo{author}{\bibfnamefont{C.~N.~J.} \bibnamefont{Wagner}},
  \emph{\bibinfo{title}{Liquid {M}etals}} (\bibinfo{publisher}{S. Z. Beer},
  \bibinfo{address}{Marcel Dekker, New York}, \bibinfo{year}{1972}).

\bibitem[{TAP()}]{TAPP}
\bibinfo{note}{TAPP, version 2.2 (1990), E. S. Microware, Inc., 2234 Wade
  Court, Hamilton, OH 45013.}

\bibitem[{\citenamefont{Ressler}(1997)}]{Ressler}
\bibinfo{author}{\bibfnamefont{T.}~\bibnamefont{Ressler}}, \bibinfo{journal}{J.
  Phys.} \textbf{\bibinfo{volume}{7}}, \bibinfo{pages}{C2}
  (\bibinfo{year}{1997}).

\bibitem[{\citenamefont{Stern et~al.}(1975)\citenamefont{Stern, Sayers, and
  Lytle}}]{Stern}
\bibinfo{author}{\bibfnamefont{E.~A.} \bibnamefont{Stern}},
  \bibinfo{author}{\bibfnamefont{D.~E.} \bibnamefont{Sayers}},
  \bibnamefont{and} \bibinfo{author}{\bibfnamefont{F.~W.} \bibnamefont{Lytle}},
  \bibinfo{journal}{Phys. Rev. B} \textbf{\bibinfo{volume}{11}},
  \bibinfo{pages}{4836} (\bibinfo{year}{1975}).

\bibitem[{\citenamefont{Rehr}(1991)}]{Rehr}
\bibinfo{author}{\bibfnamefont{J.~J.} \bibnamefont{Rehr}}, \bibinfo{journal}{J.
  Am. Chem. Soc.} \textbf{\bibinfo{volume}{113}}, \bibinfo{pages}{5135}
  (\bibinfo{year}{1991}).

\bibitem[{\citenamefont{Stern}(1993)}]{Stern2}
\bibinfo{author}{\bibfnamefont{E.~A.} \bibnamefont{Stern}},
  \bibinfo{journal}{Phys. Rev. B} \textbf{\bibinfo{volume}{48}},
  \bibinfo{pages}{9825} (\bibinfo{year}{1993}).

\bibitem[{\citenamefont{Hausleitner and Hafner}(1992)}]{Hausleitner}
\bibinfo{author}{\bibfnamefont{C.}~\bibnamefont{Hausleitner}} \bibnamefont{and}
  \bibinfo{author}{\bibfnamefont{J.}~\bibnamefont{Hafner}},
  \bibinfo{journal}{Phys. Rev. B} \textbf{\bibinfo{volume}{{45}}},
  \bibinfo{pages}{128} (\bibinfo{year}{1992}).

\bibitem[{\citenamefont{Fukunaga et~al.}(1984)\citenamefont{Fukunaga, Watanabe,
  and Suzuki}}]{Fukunaga}
\bibinfo{author}{\bibfnamefont{T.}~\bibnamefont{Fukunaga}},
  \bibinfo{author}{\bibfnamefont{N.}~\bibnamefont{Watanabe}}, \bibnamefont{and}
  \bibinfo{author}{\bibfnamefont{K.}~\bibnamefont{Suzuki}},
  \bibinfo{journal}{J. Non-Cryst. Solids} \textbf{\bibinfo{volume}{{61 \&
  62}}}, \bibinfo{pages}{343} (\bibinfo{year}{1984}).

\bibitem[{\citenamefont{Machado
  et~al.}(2004{\natexlab{a}})\citenamefont{Machado, de~Lima, Campos, Grandi,
  and Pizani}}]{KleberGeSe}
\bibinfo{author}{\bibfnamefont{K.~D.} \bibnamefont{Machado}},
  \bibinfo{author}{\bibfnamefont{J.~C.} \bibnamefont{de~Lima}},
  \bibinfo{author}{\bibfnamefont{C.~E.~M.} \bibnamefont{Campos}},
  \bibinfo{author}{\bibfnamefont{T.~A.} \bibnamefont{Grandi}},
  \bibnamefont{and} \bibinfo{author}{\bibfnamefont{P.~S.}
  \bibnamefont{Pizani}}, \bibinfo{journal}{J. Chem. Phys.}
  \textbf{\bibinfo{volume}{120}}, \bibinfo{pages}{329}
  (\bibinfo{year}{2004}{\natexlab{a}}).

\bibitem[{\citenamefont{Machado
  et~al.}(2004{\natexlab{b}})\citenamefont{Machado, J\'ov\'ari, de~Lima,
  Campos, and Grandi}}]{KleberGaSe}
\bibinfo{author}{\bibfnamefont{K.~D.} \bibnamefont{Machado}},
  \bibinfo{author}{\bibfnamefont{P.}~\bibnamefont{J\'ov\'ari}},
  \bibinfo{author}{\bibfnamefont{J.~C.} \bibnamefont{de~Lima}},
  \bibinfo{author}{\bibfnamefont{C.~E.~M.} \bibnamefont{Campos}},
  \bibnamefont{and} \bibinfo{author}{\bibfnamefont{T.~A.}
  \bibnamefont{Grandi}}, \bibinfo{journal}{J. Phys.: Condens. Matter}
  \textbf{\bibinfo{volume}{16}}, \bibinfo{pages}{581}
  (\bibinfo{year}{2004}{\natexlab{b}}).

\bibitem[{\citenamefont{Gazzillo et~al.}(1989)\citenamefont{Gazzillo, Pastore,
  and Enzo}}]{Gazzillo}
\bibinfo{author}{\bibfnamefont{D.}~\bibnamefont{Gazzillo}},
  \bibinfo{author}{\bibfnamefont{G.}~\bibnamefont{Pastore}}, \bibnamefont{and}
  \bibinfo{author}{\bibfnamefont{S.}~\bibnamefont{Enzo}}, \bibinfo{journal}{J.
  Phys.: Condens. Matter.} \textbf{\bibinfo{volume}{{1}}},
  \bibinfo{pages}{3469} (\bibinfo{year}{1989}).

\end{thebibliography}

\end{document}